\documentclass[journal=jpcbfk,manuscript=article]{achemso}

\usepackage{chemformula} % Formula subscripts using \ch{}
\usepackage[T1]{fontenc} % Use modern font encodings
\usepackage{amsfonts}
\SectionNumbersOn

\author{Radek Erban}
\affiliation[University of Oxford]{Mathematical Institute, University of Oxford, Radcliffe Observatory Quarter, Woodstock Road, Oxford, OX2 6GG, United Kingdom}
\email{erban@maths.ox.ac.uk}
\author{Yuichi Togashi}
\affiliation[Ritsumeikan University]{Department of Bioinformatics, College of Life Sciences, Ritsumeikan University, 1-1-1 Noji-higashi, Kusatsu, Shiga 525-8577, Japan}
\email{togashi@fc.ritsumei.ac.jp}
\alsoaffiliation[RIKEN]
{RIKEN Center for Biosystems Dynamics Research, 2-2-3 Minatojima-minamimachi, Chuo-ku, Kobe, Hyogo 650-0047, Japan}

\title[An \textsf{achemso} demo]
  {Asymmetric periodic boundary conditions for molecular dynamics and coarse-grained simulations of nucleic acids\footnote{This work was supported by the Engineering and Physical Sciences Research Council [EPSRC grant number EP/V047469/1] and the Japan Society for the Promotion of Science  [JSPS KAKENHI grant number JP18KK0388].}}

\abbreviations{IR,NMR,UV}
\keywords{American Chemical Society, \LaTeX}

\begin{document}

%\begin{tocentry}
%
%Some journals require a graphical entry for the Table of Contents.
%This should be laid out ``print ready'' so that the sizing of the
%Inside the \texttt{tocentry} environment, the font used is Helvetica
%8\,pt, as required by \emph{Journal of the American Chemical
%The surrounding frame is 9\,cm by 3.5\,cm, which is the maximum
%permitted for  \emph{Journal of the American Chemical Society}
%graphical table of content entries. The box will not resize if the
%content is too big: instead it will overflow the edge of the box.
%
%This box and the associated title will always be printed on a
%separate page at the end of the document.
%
%\end{tocentry}

\begin{abstract}
\noindent
Periodic boundary conditions are commonly applied in molecular dynamics simulations in the microcanonical (NVE), canonical (NVT) and isothermal-isobaric (NpT) ensembles. In their simplest application, a biological system of interest is placed in the middle of a solvation box, which is chosen `sufficiently large' to minimize any numerical artefacts associated with the periodic boundary conditions. This practical approach brings limitations to the size of biological systems that can be simulated. Here, we study simulations of effectively infinitely-long nucleic acids, which are solvated in the directions perpendicular to the polymer chain, while periodic boundary conditions are also applied along the polymer chain. We study the effects of these {\it asymmetric periodic boundary conditions } (APBC) on the simulated results, including the mechanical properties of biopolymers and the properties of the surrounding solvent. To get some further insights into the advantages of using the APBC, a coarse-grained worm-like chain model is first studied, illustrating how the persistence length can be extracted from local properties of the polymer chain, which are less affected by the APBC than some global averages. This is followed by all-atom molecular dynamics simulations of DNA in ionic solutions, where we use the APBC to investigate sequence-dependent properties of DNA molecules and properties of the surrounding solvent.
\end{abstract}

\section{Introduction}

The structure and function of DNA depends on its nucleotide sequence and on the properties of the surrounding solvent~\cite{Vologodskii:2015:BD}. Since DNA is negatively charged, concentrations of ions are perturbed from their bulk values in the region close to DNA. The resulting `ion atmosphere' has been studied using ion counting experiments~\cite{Jacobson:2016:CIS}. From the theoretical point of view, all-atom molecular dynamics (MD) simulations can be applied to provide detailed insights into DNA, ions and water interactions~\cite{Mocci:2012:INA}. For example, the effect of mobile counterions, Na$^+$ and K$^+$ on a DNA oligomer was studied by V\'arnai and Zakrzewska~\cite{Varnai:2004:DIC}, who used periodic boundary conditions for MD simulations of the solvated DNA oligomer at constant temperature and pressure and studied the counterion distribution around the DNA structure. 

However, the applicability of all-atom MD studies is limited to relatively small systems. To simulate larger systems, several coarse-grained approaches have been developed in the literature. In adaptive resolution studies~\cite{Zavadlav:2015:ARS,Zavadlav:2018:ARS}, DNA and its immediate neighbourhood are simulated using the full atomistic resolution, while a coarse-grained description is used to describe the solvent molecules which are far away from DNA. Solvent can also be treated implicitly in far away regions~\cite{Zavadlav:2018:ARS}. To model even larger systems, the DNA molecule itself can be described by coarse-grained models~\cite{Dans:2016:MSD,Korolev:2016:MCM,Poppleton:2023:OCS,Sengar:2021:POM}. Examples vary from models using several coarse-grained sites per nucleotide~\cite{Kovaleva:2017:SCD} to Brownian dynamics simulations~\cite{Rolls:2017:VRR,Maffeo:2020:MMM}. Using a systematic `bottom-up approach', the interaction potential between coarse-grained sites can be derived from the underlying atomistic force field, with results dependent on the microscopic force-field used~\cite{Minhas:2020:MDF}.

\begin{figure}[t]
\leftline{(a) \hskip 11 cm (b)}
\leftline{
\hskip 2mm
\includegraphics[height=7cm]{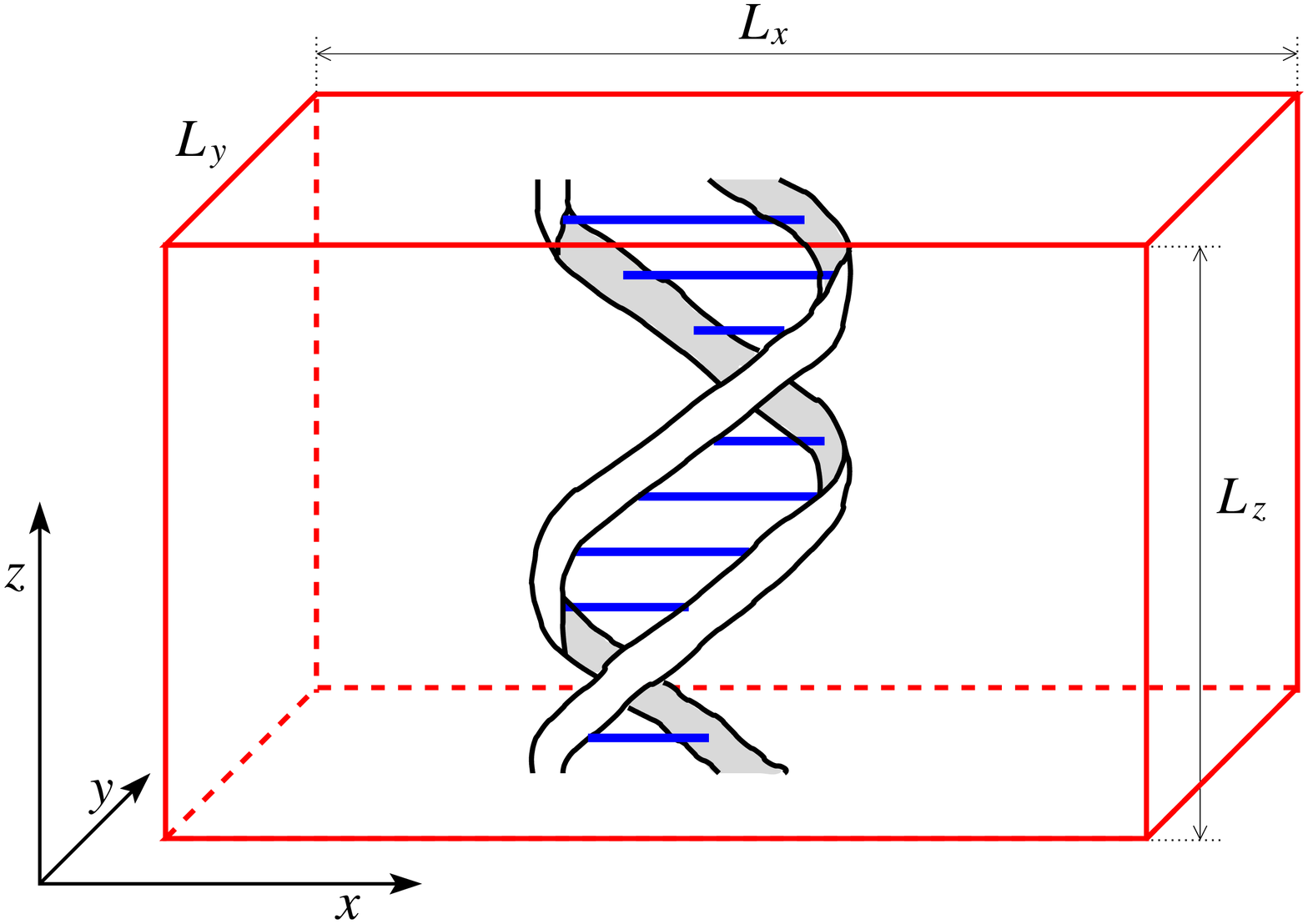}
\hskip 1.2cm 
\includegraphics[height=7cm]{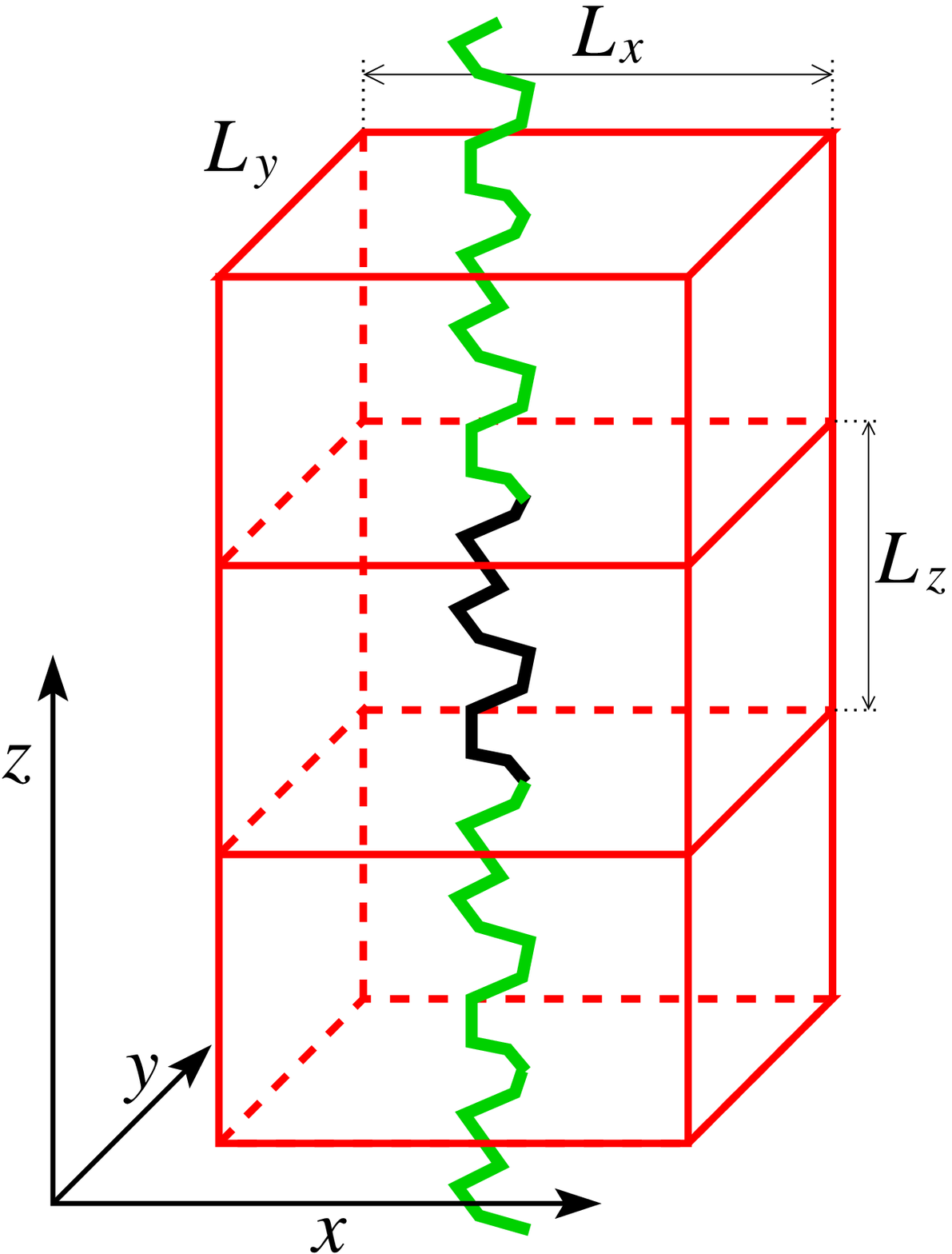}}
\caption{(a) {\it Schematic of the simulation domain $(\ref{omegadomain})$ for the case with $10$ base pairs of DNA, i.e. for $n=1.$} (b) {\it Discrete worm-like chain segment in an APBC simulation is denoted by the black line. Its periodic image copies by the green line.}}
\label{figure1}
\end{figure}

The properties of ions in bulk water can be studied using all-atom MD simulations~\cite{Lee:1996:MDS,Erban:2016:CAM}, which provide `bottom-up' estimates of the values of diffusion constants of ions. Some biological processes include transport of ions across relatively large distances which are out of reach to all-atom MD simulations. Brownian dynamics descriptions of ions are instead used for modelling such systems~\cite{Dobramysl:2016:PMM,Erban:2020:SMR}. While the transport of ions in bulk water can be described on a sufficiently long time scale as a standard Brownian motion, more detailed coarse-grained stochastic models of ions in bulk water have to be used at time scales studied by MD simulations~\cite{Erban:2016:CAM}. Coarse-grained stochastic models of ions can be written as systems of stochastic differential equations or by the generalized Langevin equation~\cite{Erban:2016:CAM,Erban:2020:CMD}. To parametrize such models, detailed all-atom MD simulations of `long' chains of nucleic acids can be used, but this can be computationally intensive. In this paper, we investigate simulations of effectively infinitely-long DNA by applying {\it asymmetric periodic boundary conditions} (APBC) in the cuboid computational domain
\begin{equation}
\Omega = [0,L_x] \times [0,L_y] \times [0,L_z].
\label{omegadomain}
\end{equation}
The main idea behind the APBC is that DNA is periodic with the period of 10~base pairs, i.e. the APBC will allow us to use $10 \, n$ base pairs of DNA in domain $\Omega$, where $n \in {\mathbb N}$ is an integer denoting the number of helical pitches.  A schematic of our simulation domain is presented in Figure~\ref{figure1}(a) for the case of the simulation with 10~base pairs, i.e. for $n=1$. The DNA molecule is positioned parallel to the $z$-axis and we use periodic boundary conditions in the $z$-direction. Such a periodic boundary condition in $z$-direction is less common in all-atom MD simulation studies, where the biomolecule of interest is often placed in the middle of the computational domain and it is solvated on all its sides by a layer of water molecules separating the biomolecule from the domain boundary~\cite{Varnai:2004:DIC,Kameda:2021:met}. 

Considering the projection into the $xy$-plane, the DNA molecule is positioned in the middle of the simulated domain. In particular, the DNA molecule is separated by the layer of water molecules from the boundaries of the simulated domain in both $x$-direction and $y$-direction. While we use periodic boundary conditions in all three directions, there is an asymmetry (highlighted in our terminology APBC): a modeller has a relative freedom to choose the values of $L_x$ and $L_y$ in the computational domain defined by~(\ref{omegadomain}), while the value of $L_z$ is dictated by the properties of the simulated biomolecule. The imposed DNA periodicity fixes the helical twist of the DNA molecule with the simulation box size $L_z$ chosen such that it exactly corresponds to $n$ helical pitches. However, considering simulations at isothermal-isobaric (NpT) ensemble, standard isotropic barostats introduce fluctuations in the domain size leading to changes in $L_z$ as well. To fix $L_z$, an asymmetric barostat is used in Section~\ref{secallatomMD} of this paper.

The APBC has been used in previous studies~\cite{Lyubartsev:1998:MDS,Korolev:2003:MDS,Zavadlav:2015:ARS} to mimic an infinitely long DNA molecule. Except of the asymmetry between the $z$-direction and $x$-direction (resp. $y$-direction), the APBC can lead to a relatively standard all-atom MD set up, with the domain periodic in all three directions, which was previously used to explore the ion atmosphere around the DNA~\cite{Lyubartsev:1998:MDS,Korolev:2003:MDS}. However, it is more challenging to use the APBC to study mechanical properties of biopolymers, as we will first illustrate in Section~\ref{WLCmodel} by considering a discrete worm-like chain model. This is followed by all-atom MD simulations of DNA in Section~\ref{secallatomMD}, where we present the use of APBC to investigate mechanical properties of the DNA  and the properties of the surrounding solvent.

\section{Worm-like chain model}

\label{WLCmodel}

\begin{figure}[t]
\leftline{(a) \hskip 8 cm (b)}
\centerline{
\hskip 5mm
\includegraphics[height=7cm]{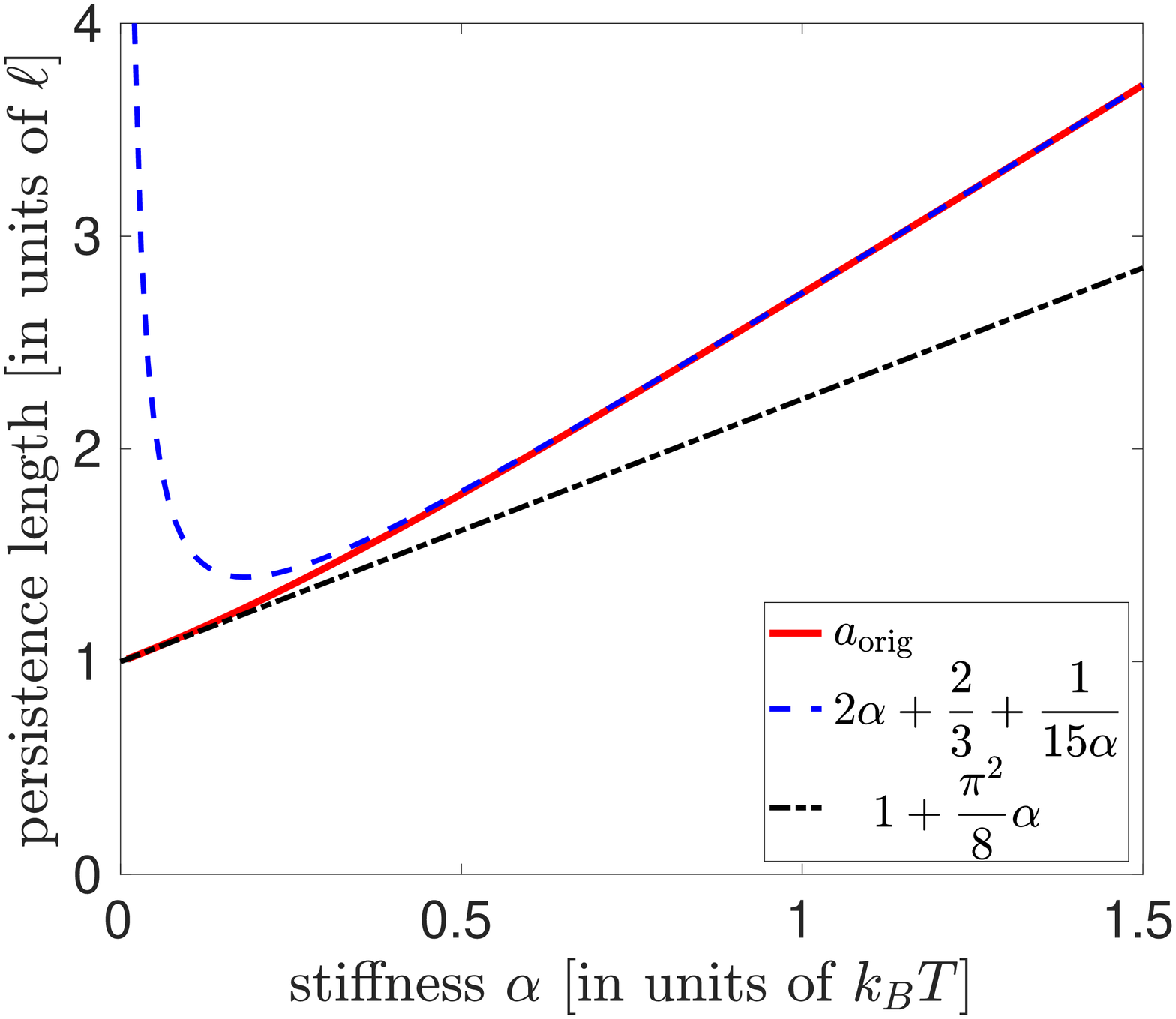}\hskip 2mm \includegraphics[height=7cm]{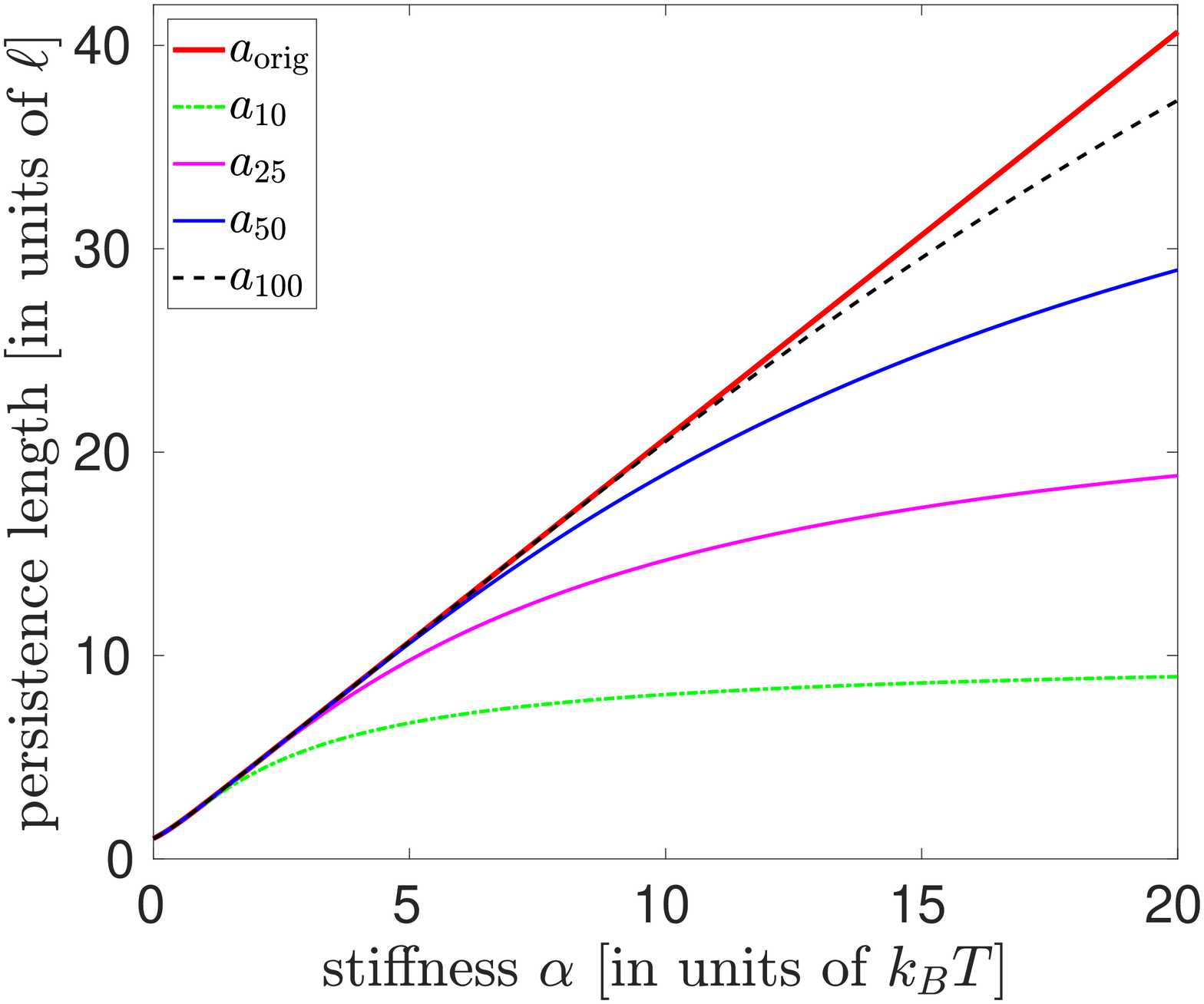} \hskip 2mm}
\caption{(a) {\it Plot of persistence length $a_{{\mathrm orig}}$, given by~$(\ref{defperslengthorig})$,
as a function of the stiffness parameter $\alpha$, together with asymptotic results~$(\ref{persform})$ and $(\ref{persformsmall}).$ The dimensionless parameter $\alpha$ can be viewed to express energy in units $k_B T$, while all persistence lengths are plotted in units of the segment lentgh, $\ell.$} (b) {\it Plot of persistence length $a_{{\mathrm orig}}$, given by~$(\ref{defperslengthorig})$, and persistence lengths $a_j$, given by~$(\ref{ajcosth})$, for $j=10,$ $20,$ $50,$ $100$, as a function of parameter $\alpha$.} \hfill\break
}
\label{figureWLC1}
\end{figure}

Let us consider the discrete worm-like chain (WLC) model where DNA consists of $N$ segments~$\mathbf{l}_i$, $i=1,2,\dots,N$, each having the same length, $\ell$. Denoting the angle between the adjacent $i$-th and $(i+1)$-th segments by~$\theta_i$, for $i=1,2,\dots,(N-1)$, the chain bending energy is
\begin{equation}
\frac{\mathrm{E}}{k_{\mathrm B} T}
=
\alpha 
\sum_{i=1}^{N-1} \theta_i^2 \, ,
\label{origenergy}
\end{equation}
where $\alpha$ is a dimensionless constant.
We define the persistence length of the first $j$-th segments, for $j \le N$, 
by
\begin{equation}
 a_j
 =
\left\langle
\frac{\mathbf{l}_1}{\ell}
\cdot
\sum_{i=1}^j
\mathbf{l}_i
\right\rangle.
\label{ajdef}
\end{equation}
That is, $a_j$ is the average value of the projection of the vector connecting the end points of the first and the $j$-th segment on the direction of the first segment. Then the persistence length of the WLC model can be defined as the limit
\begin{equation}
a_{\mathrm{orig}}
=
\lim_{N \to \infty}
a_N,
\label{defperslengthorig}
\end{equation}
which effectively is the
average value of the projection of the end-to-end vector of a long chain on the direction of the first segment. The average in~(\ref{ajdef}) can be evaluated as
\begin{equation}
a_j
=
\sum_{i=1}^j
\frac{\left\langle
\mathbf{l}_1 \cdot \mathbf{l}_i
\right\rangle}{\ell}
=
\ell
\sum_{i=1}^j
\left\langle
\cos(\theta)
\right\rangle^{i-1}
=
\ell
\,
\frac{1 - \left\langle
\cos(\theta)
\right\rangle^j}{1-\left\langle
\cos(\theta)
\right\rangle}
\, ,
\label{ajcosth}
\end{equation}
where the average $\left\langle
\cos(\theta)
\right\rangle$ is given by
\begin{equation}
\left\langle
\cos(\theta)
\right\rangle
=
\int_0^\pi
\cos(\theta)
\,
\sin(\theta) 
\,
\exp \left[
- \alpha \, \theta^2
\right]
\,
\mbox{d}\theta
\Bigg/
\int_0^\pi
\sin(\theta) 
\,
\exp \left[
- \alpha \, \theta^2
\right]
\,
\mbox{d}\theta.
\label{costhintegral}
\end{equation}
To get formula~(\ref{costhintegral}), we note that the distribution of angles between adjacent segments
is proportional to $\sin(\theta) 
\,
\exp \left[
- \alpha \, \theta^2
\right]$. 
Using (\ref{defperslengthorig}), (\ref{ajcosth}) and (\ref{costhintegral}), we deduce
\begin{eqnarray}
a_{\mathrm{orig}}
=
\frac{\ell}{1-\left\langle
\cos(\theta)
\right\rangle}
\, 
& = &
\ell \left(
2 \alpha + \frac{2}{3} + \frac{1}{15 \alpha} + \mathcal{O} \left( \frac{1}{\alpha^2} \right)
\right),
\qquad \mbox{as} \quad \alpha \to \infty,
\label{persform}
\\
& = &
\ell
\left(
1
+
\frac{\pi^2}{8} \, \alpha
+
\mathcal{O}(\alpha^2)
\right),
\qquad \mbox{as} \quad \alpha \to 0.
\label{persformsmall}
\end{eqnarray}
In Figure~\ref{figureWLC1}(a), we present how the persistence length $a_{\mathrm{orig}}$  depends on the stiffness parameter~$\alpha$ in interval $[0,3/2]$, illustrating the accuracy of both expansions (\ref{persform}) and (\ref{persformsmall}). While (\ref{persform}) is derived in the limit $\alpha \to \infty$, it approximates the exact result well for persistence lengths satisfying $a_{\mathrm{orig}} > 2 \ell$ or equivalently for $\alpha > 0.62$.
In Figure~\ref{figureWLC1}(b), we plot the dependence of the persistence length $a_{\mathrm{orig}}$ on the stiffness parameter $\alpha$ in a larger interval $[0,20]$ together with the values of $a_j$ given by~(\ref{ajcosth}). Using the exact result for $a_{\mathrm{orig}}$ given on the left hand side of equation~(\ref{persform}), we can rewrite~(\ref{ajcosth}) as follows
\begin{equation}
a_j = a_{\mathrm{orig}}
\left(
1 - \left\langle
\cos(\theta)
\right\rangle^j
\right).
\label{ajorig}
\end{equation}
Considering the limit $\alpha \to \infty$ in~(\ref{costhintegral}), we have
$$
\left\langle
\cos(\theta)
\right\rangle
=
1
-
\frac{1}{2\alpha}
+
\frac{1}{6\alpha^2}
+
\mathcal{O}
\left(
\frac{1}{\alpha^3}
\right),
\qquad \mbox{as} \quad \alpha \to \infty,
$$
where the first three terms of the expansion on the right hand side provide an approximation of $\left\langle
\cos(\theta)
\right\rangle$
with about 5\% relative error for $\alpha>1$, and the relative error decreases as we increase $\alpha$, for example, the relative error is smaller than $1\%$ for $\alpha>2.$ Substituting this expansion for $\left\langle
\cos(\theta)
\right\rangle$ into equation~(\ref{ajorig}), we obtain that for sufficiently large values of $\alpha$, say for $\alpha>1$, we can calculate the persistence length
$a_{\mathrm{orig}}$ from $a_j$ by using the following formula
\begin{equation}
a_{\mathrm{orig}}
=
\frac{a_j}{
\displaystyle 
1
-
\left(
1-
\frac{1}{2\alpha}
+
\frac{1}{6\alpha^2}
\right)^j
}.
\label{ajorigalpha}
\end{equation}

\subsection{The dependence of persistence length on APBC}

\begin{figure}[t]
\leftline{(a) \hskip 8 cm (b)}
\centerline{
\hskip 5mm
\includegraphics[height=6cm]{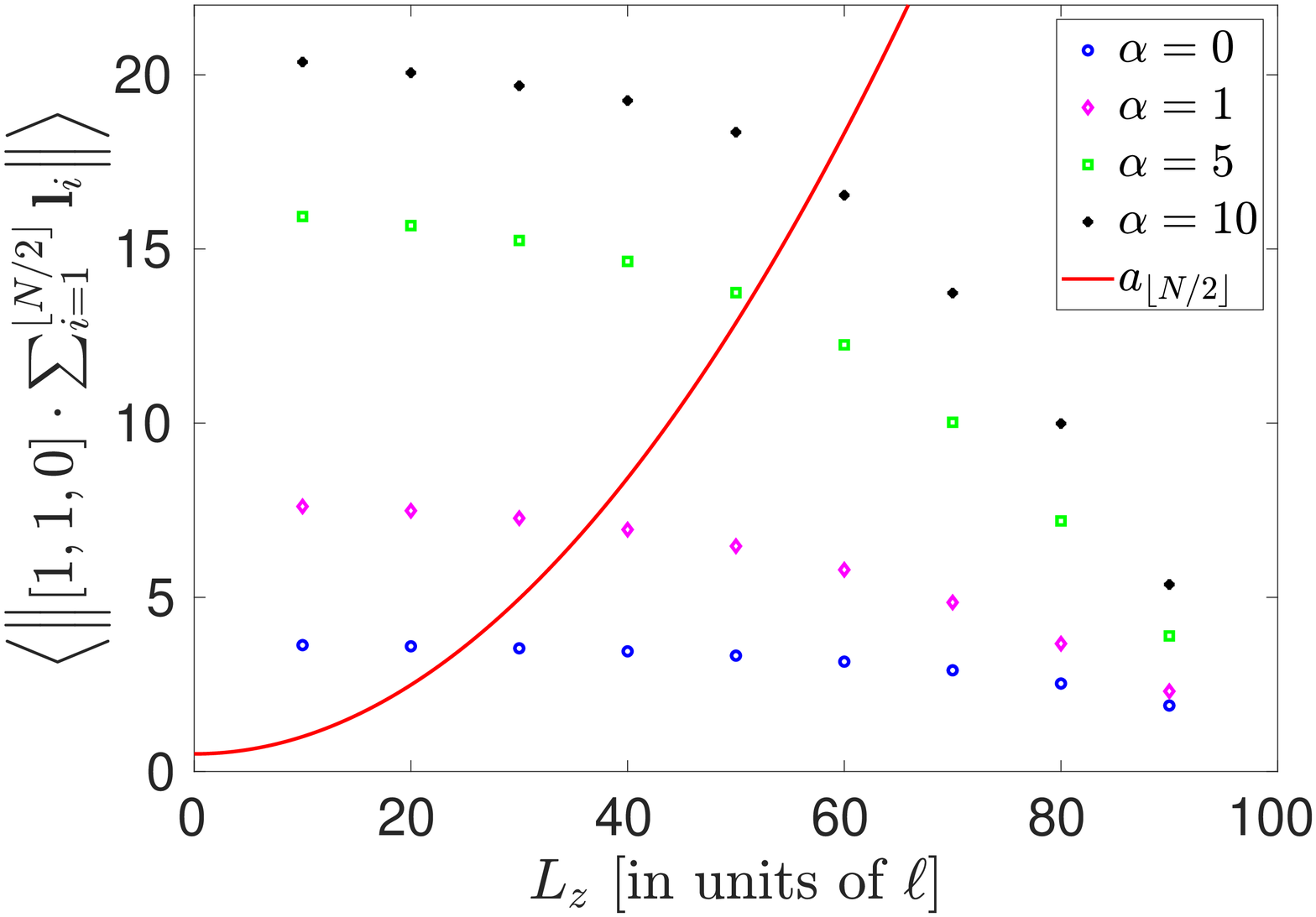}\hskip 2mm \includegraphics[height=6cm]{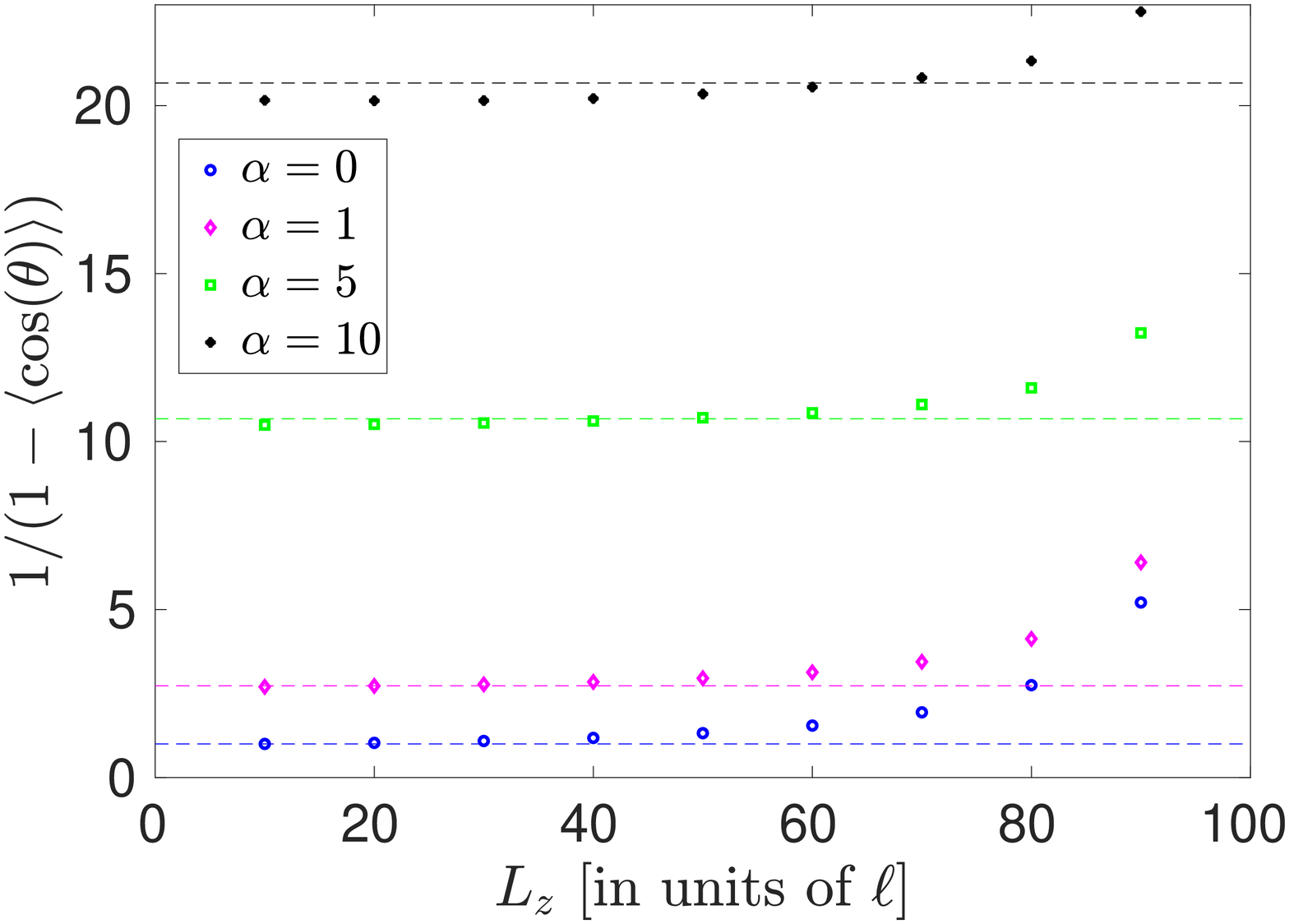} \hskip 2mm}
\caption{(a) {\it 
The average distance of the polymer middle point from the axis of the polymer, defined by~$(\ref{midwith})$, estimated from Monte Carlo simulations of the WLC model with $N=100$ segments for $L_z \in \{10, 20, \dots, 90\}$ and 
$\alpha \in \{0, 1, 5, 10 \}$. The red line shows $a_{\lfloor N/2 \rfloor} = a_{50}$ as a function of the domain length~$L_z$ (theoretical result~$(\ref{Ngenformula})$ confirmed by simulations for all considered values of $\alpha$).} \hfill \break (b) {\it The estimate of $a_{\mathrm{orig}}$ given by equation~$(\ref{aorigform})$, where $\left\langle \cos(\theta) \right\rangle$ is estimated from Monte Carlo simulations of the WLC model with $N=100$ segments for $L_z \in \{10, 20, \dots, 90\}$ and $\alpha \in \{0, 1, 5, 10 \}$. The theoretical result (without APBC, independent of $L_z$), given by equation~$(\ref{costhintegral})$, is plotted by dashed lines for each value of $\alpha.$}\hfill\break
}
\label{figureWLC2}
\end{figure}

Considering that the polymer chain is simulated in the domain~(\ref{omegadomain}) with APBC, we have an extra constraint
\begin{equation}
\sum_{i=1}^N
\mathbf{l}_i
=
[0,0,L_z] \, ,
\label{auxf1}
\end{equation} 
where $N$ denotes the number of simulated segments along the $z$-direction. As it is illustrated in Figure~\ref{figure1}(b),
such a model can be viewed as a model of an (infinitely) long polymer chain by using the periodicity
\begin{equation}
\mathbf{l}_i =  \mathbf{l}_{i \; \mathrm{MOD} \; N},
\qquad
\mbox{for}
\; i \in \mathbb{N}.
\label{auxf2}
\end{equation}
However, substituting (\ref{auxf1})--(\ref{auxf2}) into the definition of persistence length~(\ref{defperslengthorig}), we would
obtain that $a_{\mathrm{orig}} = \infty$ because the periodic boundary means that the infinitely long filament is effectively straight. Since equation~(\ref{auxf1}) postulates that the vector connecting ends of $N$ segments is fixed, we obtain the most variability in this model by looking at the behaviour of the $\lfloor N/2 \rfloor$ consecutive segments. Due to the symmetry of the problem and condition~(\ref{defperslengthorig}), the average of the vector $\sum_{i=1}^{\lfloor N/2 \rfloor} \mathbf{l}_i$ is equal to $[0,0,L_z/2]$ for any value of $\alpha$, but the deviations from this average will depend on $\alpha.$ To illustrate this, we define the average distance of the polymer middle point from the axis of the polymer by
\begin{equation}
\left\langle
\bigg\Vert
[1,1,0]
\cdot
\sum_{i=1}^{\lfloor N/2 \rfloor}
\mathbf{l}_i
\bigg\Vert
\right\rangle,
\label{midwith}
\end{equation}
that is, we calculate the (Euclidean) norm of the projection of the vector $\sum_{i=1}^{\lfloor N/2 \rfloor} \mathbf{l}_i$ on the $x$--$y$--plane. The 
average~(\ref{midwith}) is plotted in Figure~\ref{figureWLC2}(a) for different values of parameter $\alpha$ and domain length~$L_z.$ 
We observe that, for fixed value of $L_z$, the average~(\ref{midwith}) increases with the value of the stiffness parameter~$\alpha.$ Moreover, Figure~\ref{figureWLC2}(a) also shows that the value of the average~(\ref{midwith}) approaches zero as $L_z$ approaches its maximum possible value, $N \ell$. Indeed, if $L_z = N \ell$, the polymer is straight and the value of~(\ref{midwith}) is exactly equal to zero. On the other hand, if $L_z$ is smaller then we obtain a larger value of~(\ref{midwith}), especially for polymers with larger persistence length 
(i.e. for large values of $\alpha$).

On the face of it, one possible way to estimate $a_{\mathrm{orig}}$ could be to estimate $a_{\lfloor N/2 \rfloor}$ from our Monte Carlo simulations and then use formula~(\ref{ajorigalpha}) for $j=\lfloor N/2 \rfloor$. However, formula~(\ref{ajorigalpha}) has been derived for the case of the WLC model in the 3-dimensional physical space ${\mathbb R}^3$. Considering the APBC, we obtain that $a_{\lfloor N/2 \rfloor}$ is independent of $\alpha$ (see Appendix~\ref{appfjm}). We have \begin{equation}
a_{\lfloor N/2 \rfloor}
=
\ell
\left(
\frac{N-\lfloor N/2 \rfloor}{N-1}
+
\frac{L_z^2 (\lfloor N/2 \rfloor - 1)}{N (N-1) \, \ell^2}
\right),
\label{Ngenformula}
\end{equation}
which simplifies to $a_{\lfloor N/2 \rfloor} \approx (\ell/2) + L_z^2/(2 \ell N)$ for large values of $N.$
This result is also visualized in Figure~\ref{figureWLC2}(a). In particular, a better strategy to obtain the real persistence length $a_{\mathrm{orig}}$ from the APBC simulations is to estimate 
$$
\left\langle
\mathbf{l}_i
\cdot 
\mathbf{l}_{i+1}
\right\rangle
=
\left\langle
\cos(\theta)
\right\rangle
$$ 
and then use the exact result for $a_{\mathrm{orig}}$ given on the left hand side of equation~(\ref{persform}), namely
\begin{equation}
a_{\mathrm{orig}}
=
\frac{\ell}{1-\left\langle
\cos(\theta)
\right\rangle}.
\label{aorigform}
\end{equation}
The results are presented in Figure~\ref{figureWLC2}(b).

\section{APBC in all-atom MD simulations}

\label{secallatomMD}

In this section, we investigate the use of APBC in all-atom MD models of DNA. Our simulations are performed with 10--100 base pairs (bp)  of double-stranded DNA (dsDNA). Since we use the APBC, all simulations are effectively simulating (infinitely) long DNA chains. In particular, MD results with the longest simulated chain (100 bp) can be used as the `ground truth' for the presented APBC simulations with shorter 10--50 bp long DNA chains. We note that the MD simulations of relatively short 50 bp DNA segments without APBC have been previously used in the literature to estimate the DNA persistence length by using a middle section of the simulated DNA segment~\cite{Kameda:2021:met}. 

We consider 6 types of (infinitely) long DNA sequences, with repeated nucleotides, namely poly(A), poly(C), poly(AT), poly(CG), poly(AC) and poly(AG), where poly(X) means that the correspoding nucleotide sequence is periodically repeated. We note that these 6 cases correspond to all possible cases of pairs of nucleotides which are repeated infinitely many times. For example, repetitions of dinucleotides AC, CA, TG and GT all correspond to the poly(AC) case, because AC and CA are equivalent due to the periodic boundary conditions along the chain length, and TG is on the complementary strand, with GT being equivalent to TG because of the periodic boundary conditions. 

Each infinitely long sequence is modelled in our computational domain~(\ref{omegadomain}) with APBC using $N=10 \, n$ base pairs of DNA, where $n$ ranges from 1 to 10. The APBC is implemented along the $z$-direction as detailed in Appendix~\ref{appAPBC}. First, an $(N+1)$ bp long dsDNA configuration is constructed in such a way that the $(N+1)$-th base pair is equivalent to the first base pair translated to the $z$-direction. Then, a nucleotide at the 3'-end of each strand is removed and the bond to the 3'-end (removed) nucleotide is substituted with that to the first base at the 5'-end. The corresponding angles and dihedrals are added to MD structural files as detailed in Table \ref{tab:psf} in Appendix~\ref{appAPBC}. In all MD simulations, we consider domain~(\ref{omegadomain}) with $L_{x} = L_{y} = 200$~\AA{} and we vary $L_z$. In Figures~\ref{figure4},~\ref{figure5} and~\ref{figure7}, we choose $L_z$ as a multiple of $n$ (resp. $N$) with
\begin{equation}
L_{z} 
= 
3.375 N \, \mbox{\AA}
=
33.75 n \, \mbox{\AA},
\label{Lzchoiceinfig457}
\end{equation}
while we study the effect of stretching and shrinking of DNA in Figure~\ref{figure6} by using $L_z$ obtained as the 95\%, 100\% and 105\% of the value given by equation~(\ref{Lzchoiceinfig457}). All MD simulations are done in KCl solutions, with $K^{+}$ ions neutralizing the negatively charged DNA segments. We use the concentration 150 mM KCl in Figures~\ref{figure4},~\ref{figure5} and~\ref{figure6}, while we vary the concentration of KCl in Figure~\ref{figure7}.

When using APBC with polymer models, there are (locally) two important directions: parallel to the polymer chain and perpendicular to the polymer chain. We consider both of them, in Sections~\ref{secperallatomMD} and~\ref{secionathmosphere}, respectively. In Section~\ref{secperallatomMD}, we study the effects of APBC on the properties of the DNA chain, where we can make direct analogues to the results obtained for the persistence length of the WLC model in Section~\ref{WLCmodel}. This is followed by studying the characteristics of the surrounding solvent in Section~\ref{secionathmosphere}, where we investigate the ion athmosphere around DNA for different concentrations of KCl.

\subsection{Mechanical properties along the chain}

\label{secperallatomMD}

The persistence length for our (infinitely) long sequences of dinucleotides can be determined by various experimental and theoretical studies~\cite{Geggier:2010:SDD} as summarized in Appendix~\ref{appMD}. In Figure~\ref{figure4}, we present the results of all-atom MD simulations  with APBC of $N=10$ bp segments using the six cases of repeated dinucleotides. Technical details of these MD
simulations are given in Appendix~\ref{appMD}.

To analyze our MD results, we associate a unit orientation vector ${\mathbf h}_i$ with each base pair, i.e. $i=1,2,\dots,N$, where $N=10n$ is the total number of simulated base pairs. Denoting the angle between the $i$-th and $(i+j)$-th base pair as $\phi_j$, we have $\cos(\phi_j) = {\mathbf h}_i \cdot {\mathbf h}_{i+j}$, which we calculate for all $i=1,2,\dots,N$. Averaging the calculated results over all possible values of $i$, we have 
\begin{equation}
\left\langle
\cos(\phi_j)
\right\rangle
=
\left\langle
{\mathbf h}_i \cdot {\mathbf h}_{i+j}
\right\rangle,
\label{cosphij}
\end{equation}
where the accuracy of this average is further improved by calculating it as a time average over long MD time series. More precisely, we calculate three independent time series of length~10~ns and sample our results every 10~ps, disregarding the beginning of each simulation as the time required to equilibrate the system, see Appendix~\ref{appMD} for more details. Considering $N=10$ (i.e. $n=1$), we plot the averages~(\ref{cosphij}) in Figure~\ref{figure4}(a) for values $j=1,2,3,4,5$.
\begin{figure}[t]
\leftline{(a) \hskip 8 cm (b)}
\centerline{
\hskip 5mm
\includegraphics[height=6cm]{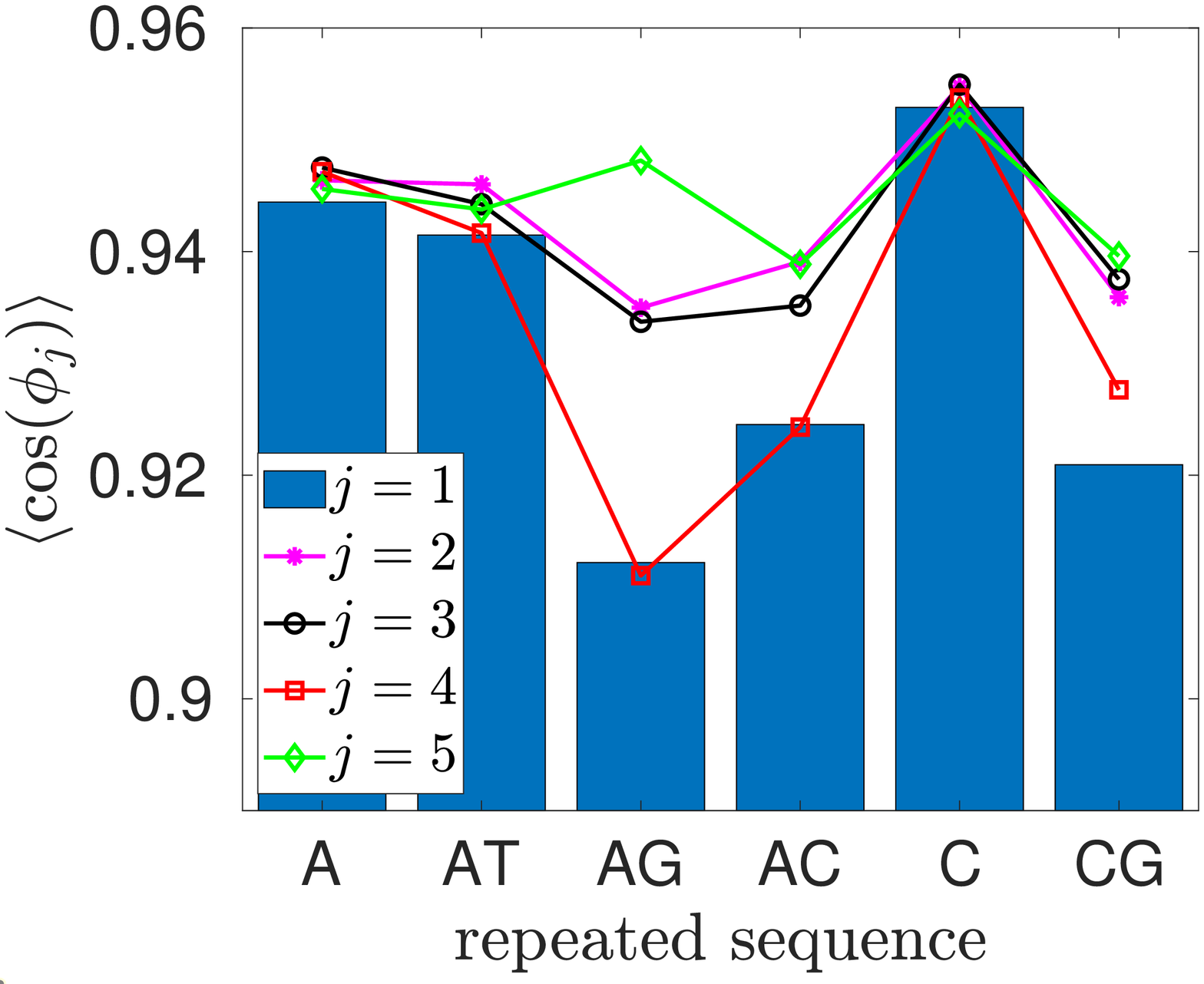}\hskip 2mm \includegraphics[height=6cm]{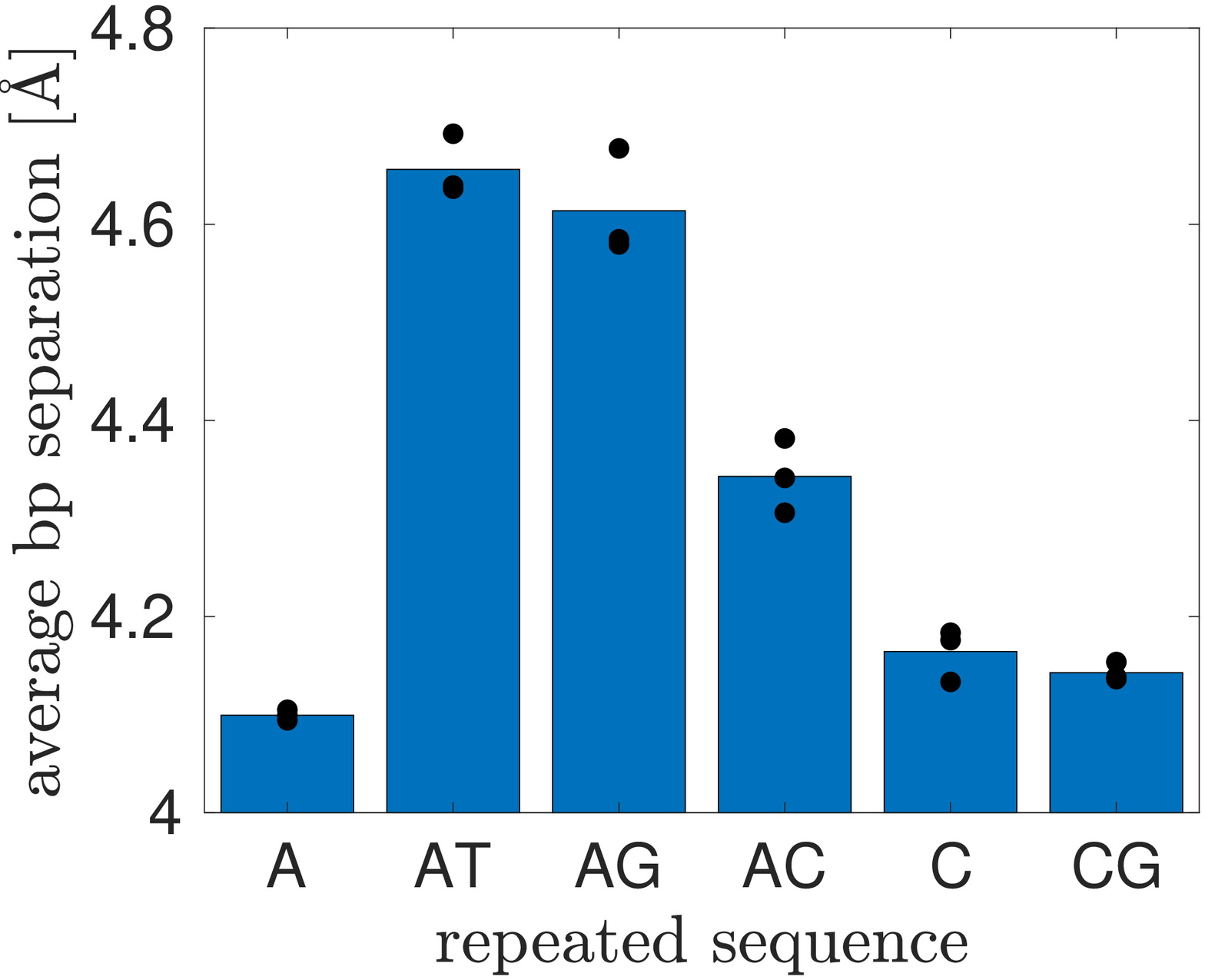} \hskip 2mm}
\caption{{\it The results of all-atom MD simulations of DNA chains with APBC.} \hfill\break (a) {\it The average~$(\ref{cosphij})$ for each of the $6$ considered sequences of repeated nucleotides is calculated using three independent MD simulations. \hfill\break
} (b) {\it The average separation between the base pairs calculated using three independent MD simulations (blue bars). The results for each individual realization is plotted as a black dot.
}}
\label{figure4}
\end{figure}%
We note that 
$$
\left\langle
\cos(\phi_j)
\right\rangle
=
\left\langle
\cos(\phi_{N-j}
\right\rangle
=
\left\langle
\cos(\phi_{10n-j}
\right\rangle,
$$
because we use APBC. In particular, the values of the averages~(\ref{cosphij}) for $j=6,7,\dots$ are already represented
in Figure~\ref{figure4}(a) by the corresponding values for $j=1,2,3,4,5$. In Figure~\ref{figure4}, we observe that the results are clearly sequence dependent for $j=1$, with the $j=4$ case providing the best match to the $j=1$ case. On the other hand, the results are less sequence dependent for $j=2$ or $j=5.$ Given the APBC, there is no variation for $j=N=10$ as we have already observed for the WLC model, because of the constraint~(\ref{auxf1}). In Figure~\ref{figure4}(b), we present the average separation between the subsequent base pairs for each of the studied case. 

\begin{figure}[t]
\leftline{(a) \hskip 8 cm (b)}
\centerline{
\hskip 5mm
\includegraphics[height=6cm]{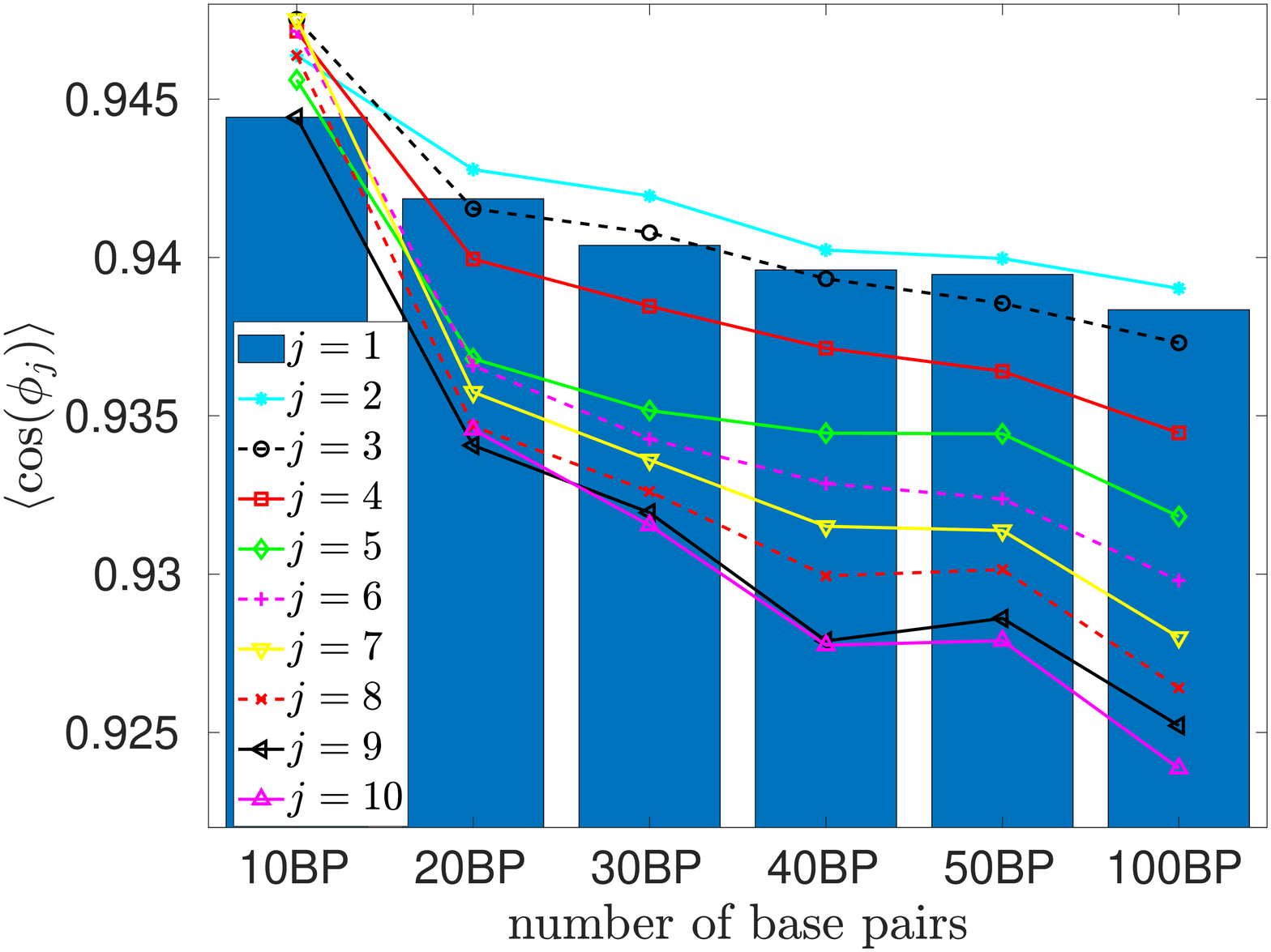}\hskip 2mm \includegraphics[height=6cm]{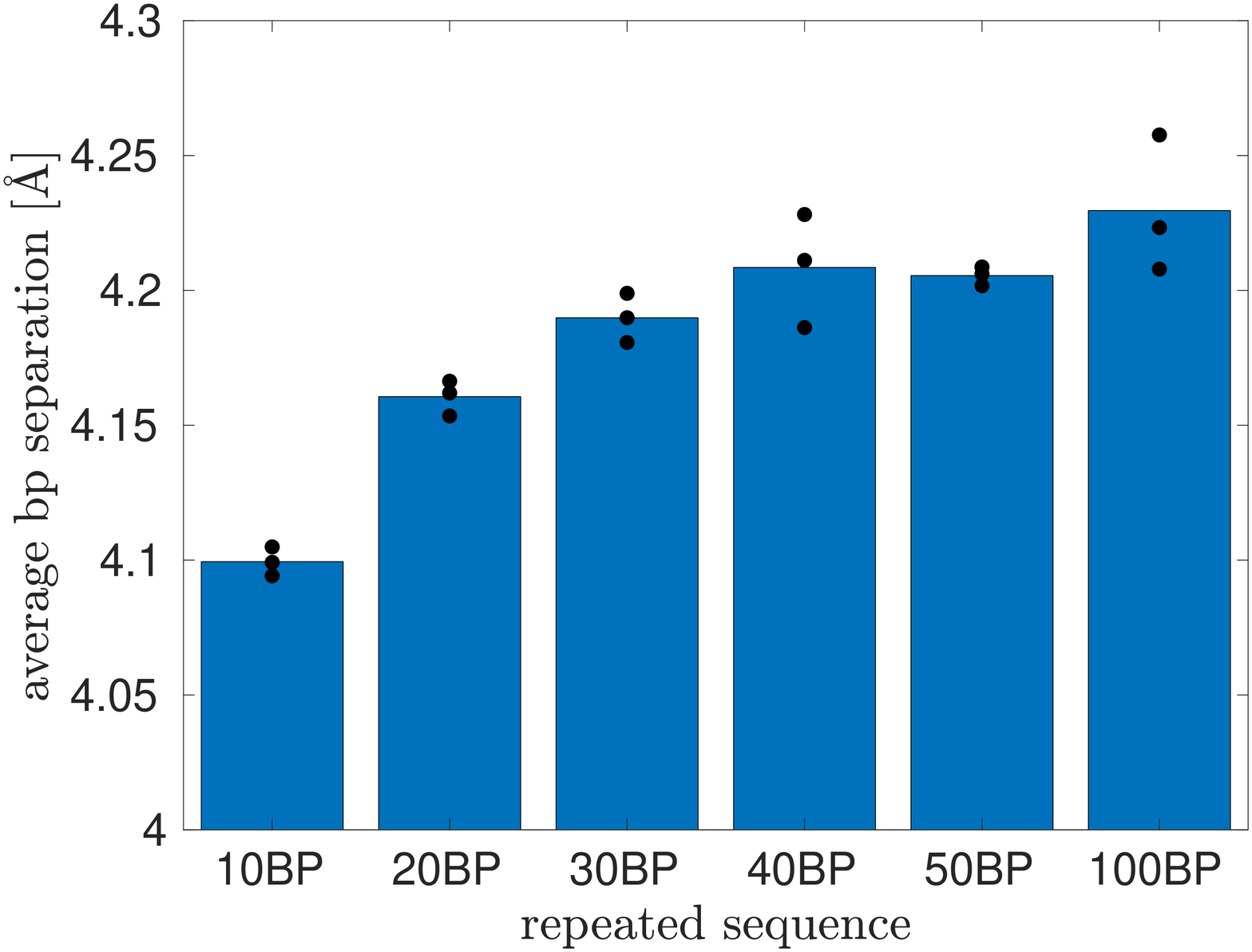} \hskip 2mm}
\caption{{\it The results of all-atom MD simulations with APBC using the poly(A) DNA chain with $N$ in the range $10$--$100$ bp.} \hfill\break (a) {\it The average~$(\ref{cosphij})$ for each of the $6$ values of $N$ considered is calculated using three independent MD simulations. The results are presented for $j=1,2,\dots,10$ and $n=1,2,3,4,5,10$. 
} \hfill \break (b) {\it The average separation between the base pairs calculated using three independent MD simulations (blue bars). The results for each individual realization is plotted as a black dot. \hfill\break
}}
\label{figure5}
\end{figure}

Our MD simulations in Figure~\ref{figure4} use the smallest possible value of $N$ (corresponding to $n=1$), while one can expect that the results of all-atom MD simulations should be less influenced by the APBC for larger values of $n$ (in theory, the APBC-induced errors should decrease to zero in the limit $n \to \infty$). To investigate this further, we study the dependence of our results on $n$ for the poly(A) case in Figure~\ref{figure5}(a). We use three independent MD simulations for $n=1, 2, 3, 4, 5, 10$ corresponding to simulations with $N$ ranging from 10bp to 100bp. In each case, we plot the averages~(\ref{cosphij}) for $j=1,2,\dots,10$. We note that this average is trivially equal to 1 in the case $j=10$ for $N=10$ bp (because the first and the eleventh base pairs are identical for $N=10$ bp), so we omit this artificial value from our plot for 10 bp in Figure~\ref{figure5}(a). We observe that the results for $n=1,2,3,4,5$ are matching some trends of the results for 100 bp. In particular, we can make similar conclusions as in Section~\ref{WLCmodel} that the local properties (smaller values of $j$) are less influenced by using APBC than the averages estimated over the whole simulated polymer length (for $j$ comparable to~$N$). 

\begin{figure}[t]
\leftline{\hskip 2mm (a) \hskip 3.7 cm (b) \hskip 3.8cm (c)}
\centerline{
\hskip 5mm
\includegraphics[height=6cm]{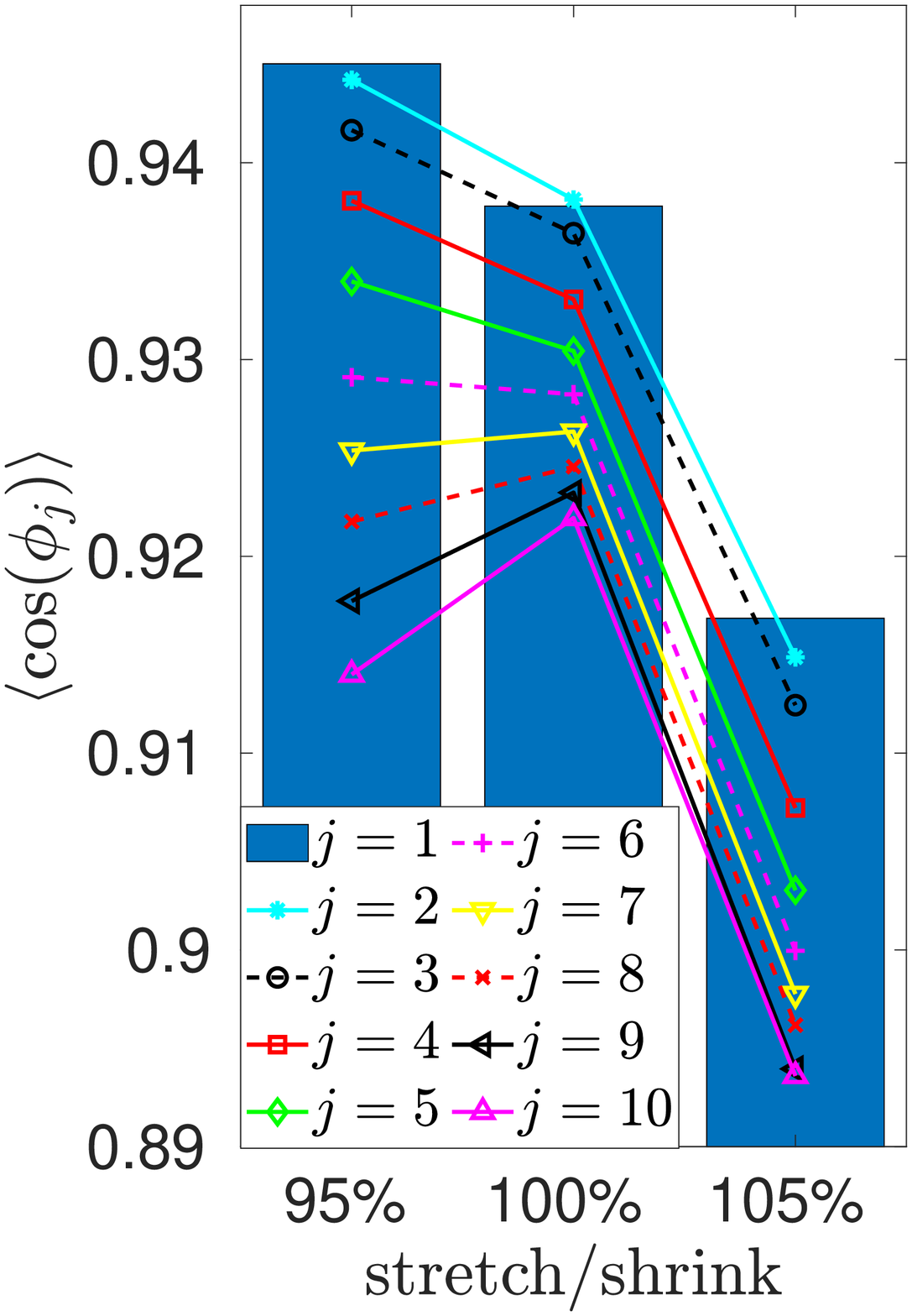}\hskip 2mm \includegraphics[height=6cm]{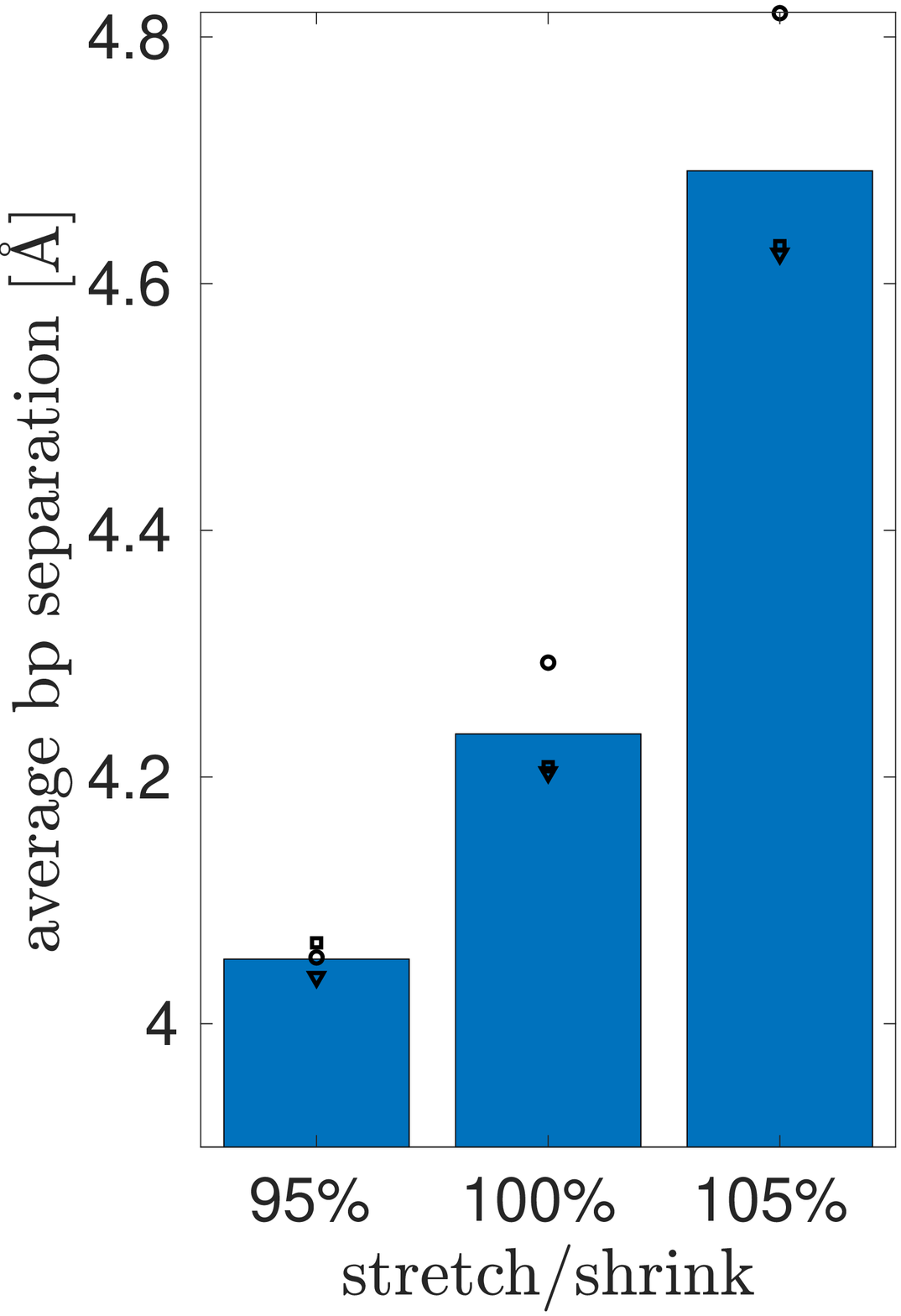} \hskip 2mm \includegraphics[height=6cm]{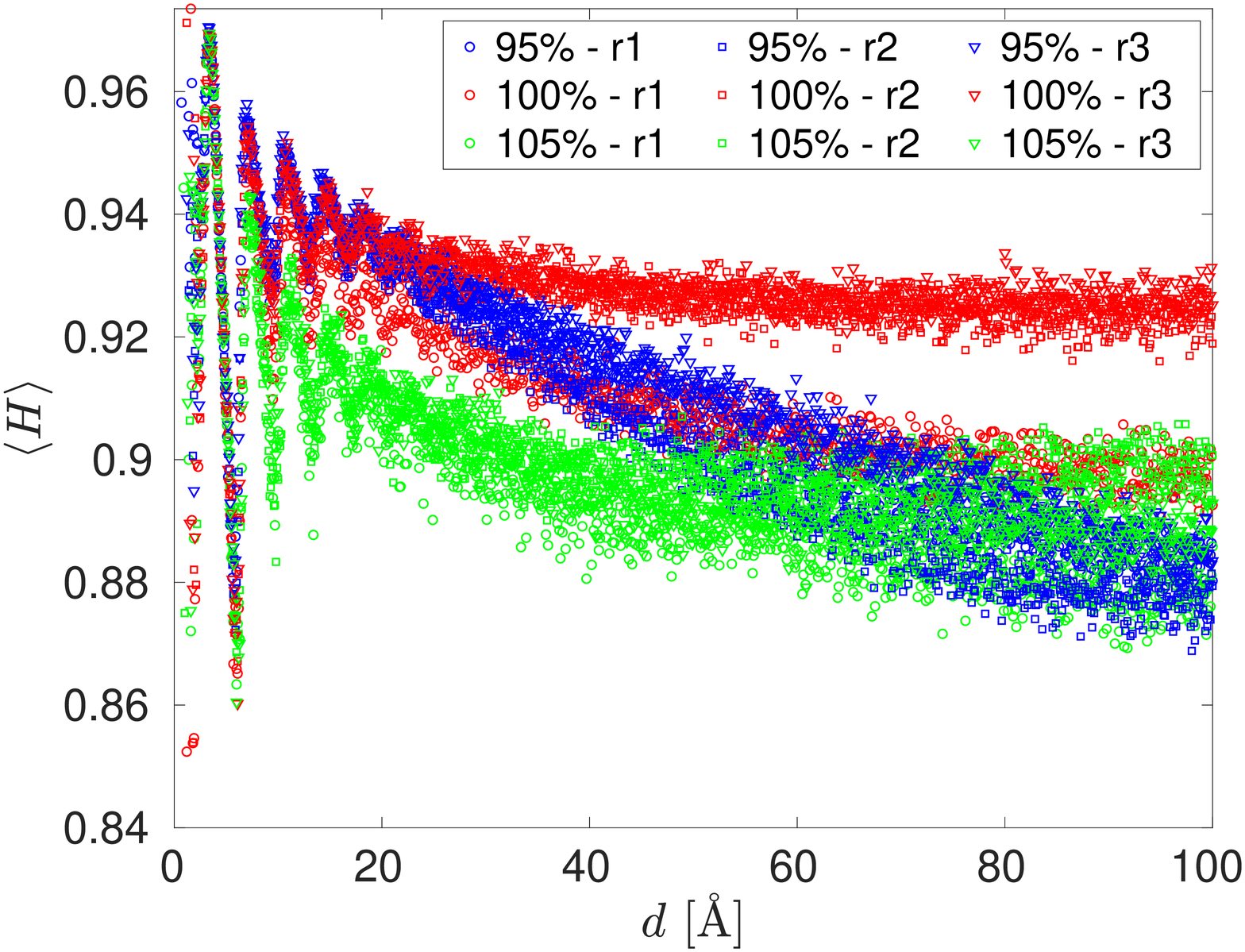} \hskip 2mm}
\caption{{\it The results of MD simulations of $100$ bp poly(A) dsDNA with APBC that use the values of $L_z$ given by equation~$(\ref{lzvalues})$. We average over three independent MD time series for each of the presented case.} \hfill\break (a) {\it The average~$(\ref{cosphij})$ for $j=1,2,\dots,10$. 
} \hfill \break (b) {\it The average separation of base pairs (blue bars). Dots include the results for individual MD realizations (i.e. we have averaged over the dots to calculate blue bars).} \hfill\break
(c) {\it The average $\langle H \rangle$ as a function of distance $d$. We present results for the $95$\% (blue), $100$\% (red) and $105$\% (green) cases using different colours. Different symbols (circle, square, triangle of the same colour) denote data points calculated by different MD realizations.} \hfill\break}
\label{figure6}
\end{figure}

In Figure~\ref{figureWLC2}, we have considered the WLC model with $N=100$ segments while varying the domain length $L_z$. In Figure~\ref{figure6}, we present the results of a similar study using all-atom MD simulations with $N=100$ bp. The middle bars in Figure~\ref{figure6}(a) and Figure~\ref{figure6}(b) correspond to the results of the poly(A) case with 100 bp which has already been included in Figure~\ref{figure5}(a). Using equation~(\ref{Lzchoiceinfig457}), this corresponds to $L_z = 337.5$~\AA. The other simulations correspond to the same set up where we either extend or shrink the value of $L_z$ by $5$\%, i.e. we use the values of $L_z$ given as
\begin{equation}
L_z = 320.625 \,\mbox{\AA}, \qquad 
L_z = 337.5 \,\mbox{\AA}, \qquad \mbox{and} \qquad
L_z = 354.375 \,\mbox{\AA}.
\label{lzvalues}    
\end{equation}
In Figure~\ref{figure6}(b), we observe that the average separation between base pairs increases as we increase $L_z$. On the other hand, the behaviour of averages~(\ref{cosphij}) is less monotonic as we stretch or shrink the DNA chain, see Figure~\ref{figure6}(a). Another way to visualize the results of all atom MD simulations
is to consider the average~(\ref{cosphij}) as a function of the distance between the base pairs~\cite{Kameda:2021:met}, which is visualized as function
$\langle H \rangle$ in Figure~\ref{figure6}(c). To calculate $\langle H \rangle$, we average $\left\langle
{\mathbf h}_i \cdot {\mathbf h}_{j} \right\rangle$ over all pairs $i$ and $j$ such that the corresponding base pairs are the distance $d$ apart. We present this average, $\langle H \rangle$, as a function of the distance $d$ in Figure~\ref{figure6}(c). The rate of decay of function $\langle H \rangle$ with distance $d$ can be used as an alternative way to define and estimate the persistence length from MD simulations~\cite{Kameda:2021:met}.

\subsection{Ion atmosphere}

\label{secionathmosphere}

\begin{figure}[t]
\leftline{(a) \hskip 8 cm (b)}
\centerline{
\hskip 5mm
\includegraphics[height=6cm]{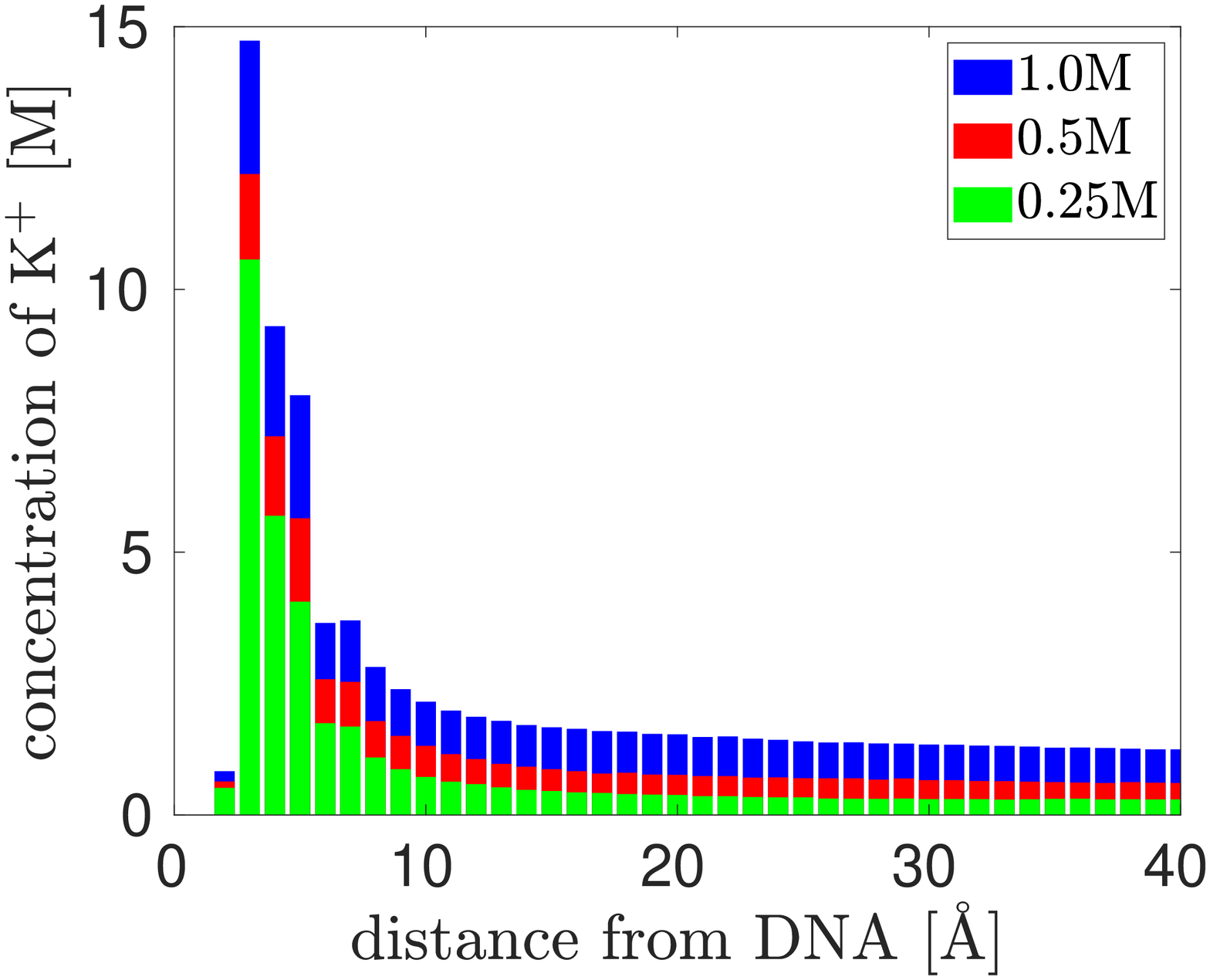}\hskip 2mm \includegraphics[height=6cm]{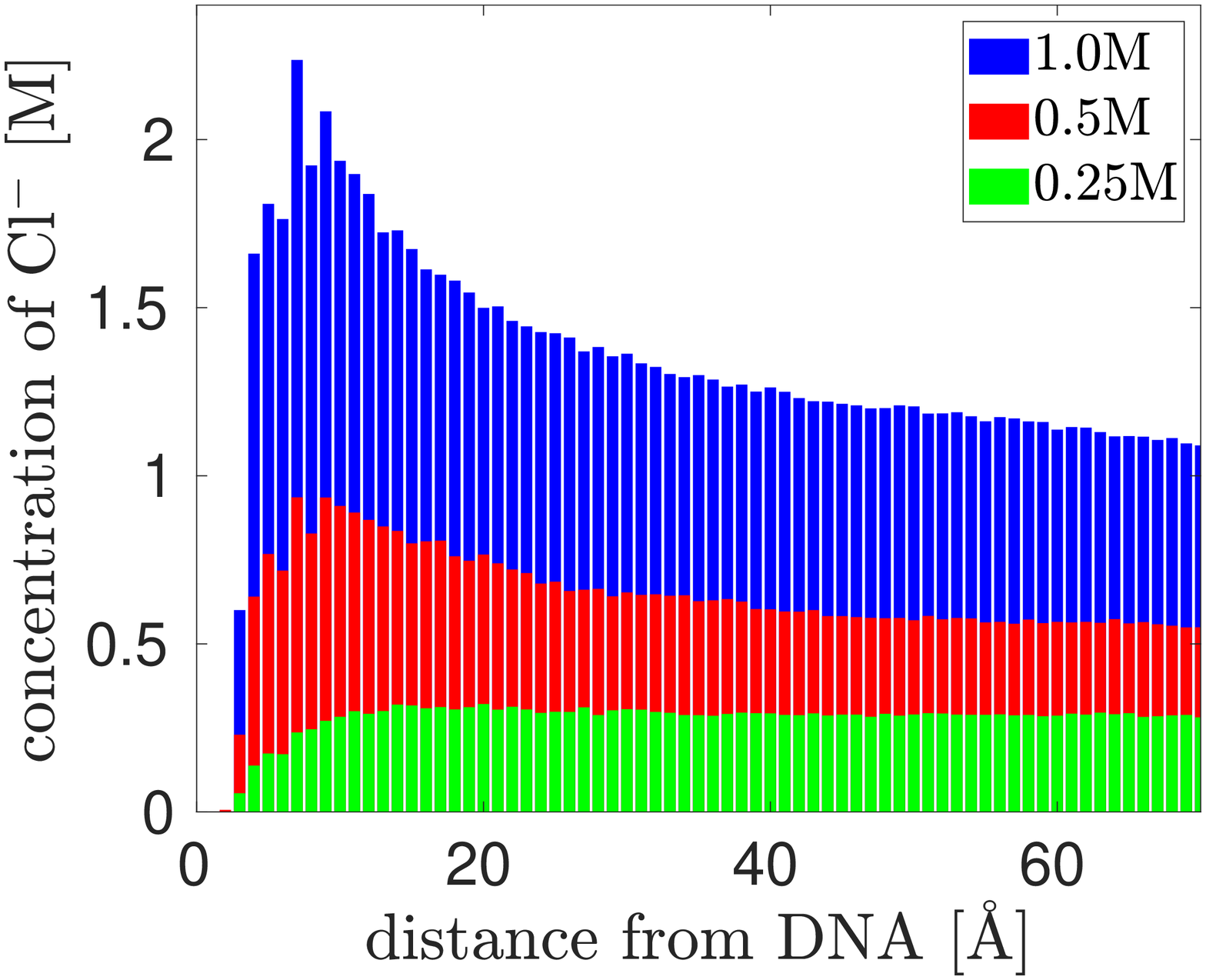} \hskip 2mm}
\caption{{\it The results of all-atom MD simulations with APBC using the poly(A) DNA chain with $N=10$ bp and three different concentrations of {\rm KCl} in the bulk.} \hfill\break(a) {\it The concentration of {\rm K}$^+$ ions given by~$(\ref{straightcyl})$ as a function of the distance from DNA.
} \hfill \break (b) {\it The concentration of {\rm Cl}$^-$ ions given by~$(\ref{straightcyl})$ as a function of the distance from DNA. \hfill\break
}}
\label{figure7}
\end{figure}

The APBC are useful for investigating solvent properties in the direction perpendicular to the polymer chain. In Figure~\ref{figure7}, we present the results of such a study, calculating the radial distribution of K$^+$ and Cl$^-$ ions. We use three different concentrations of KCl, namely~0.25M, 0.5M and 1M. In each case, we use $n=1,$ i.e. we use the APBC with 10 bp of poly(A) dsDNA. The results are calculated by averaging over four independent MD time series, each calculated for 10 ns. After the initial transient (of 1 ns) and at equidistant time intervals of 10 ps, we calculate the distance of each ion from the nearest atom of DNA, so our raw data are given in terms of the histograms
\begin{eqnarray*}
N_{\mbox{\scriptsize K}^+}(r, \Delta r)
&=&
\big[
\mbox{number of ions with the distance from DNA in interval} \; (r,r+\Delta r)
\big],
\\
N_{\mbox{\scriptsize Cl}^-}(r, \Delta r)
&=&
\big[
\mbox{number of ions with the distance from DNA in interval} \; (r,r+\Delta r)
\big].
\end{eqnarray*}
To get the radial distribution function, these numbers have to be divided by the volume, $V(r,\Delta r)$, giving the volume of all points which have their distance from the DNA in the interval $(r,r+\Delta r)$.
Then the radial distribution of K$^+$ ions and Cl$^-$ ions is defined by 
\begin{equation}
\varrho_{\mbox{\scriptsize K}^+}(r)
=
\lim_{\Delta r \to \infty}
\frac{N_{\mbox{\scriptsize K}^+}(r, \Delta r)}
{V(r,\Delta r)},
\qquad
\mbox{and}
\qquad
\varrho_{\mbox{\scriptsize Cl}^-}(r)
=
\lim_{\Delta r \to \infty}
\frac{N_{\mbox{\scriptsize Cl}^-}(r, \Delta r)}
{V(r,\Delta r)},
\label{limitdr}
\end{equation}
where $r$ is the distance from the DNA. To calculate Figure~\ref{figure7}, we approximate the limit in equation~(\ref{limitdr}) by choosing (relatively small) value $\Delta r = 1$~\AA{} and we approximate the DNA as a straight line (or equivalently as a straight cylinder) in the $z$-direction, i.e. $
V(r,\Delta r)
=
2 \pi \, r \, \Delta r \, L_z,
$
giving
\begin{equation}
\varrho_{\mbox{\scriptsize K}^+}(r)
\approx
\frac{N_{\mbox{\scriptsize K}^+}(r, \Delta r)}
{2 \pi \, r \, \Delta r \, L_z},
\qquad
\varrho_{\mbox{\scriptsize Cl}^-}(r)
\approx
\frac{N_{\mbox{\scriptsize K}^+}(r, \Delta r)}
{2 \pi \, r \, \Delta r \, L_z}.
\label{straightcyl}
\end{equation}
Formulas (\ref{straightcyl}) are visualized in 
Figure~\ref{figure7} as histograms.

\section{Discussion}

Using MD simulations at constant pressure and temperature, we can solvate the DNA with water and ions, fixing the concentration of ions in the bulk. In Section~\ref{secionathmosphere}, we have presented illustrative results of such all-atom MD investigations with APBC. Such simulations can also be used to estimate other solvent properties, for example, the moments of force distributions on ions, which can be used for parametrizing coarse-grained stochastic models of ions used in multiscale and multi-resolution simulations~\cite{Erban:2016:CAM,Erban:2020:CMD}. The APBC simulations can also be coupled with coarse-grained models of water to design adaptive resolution simulation techniques~\cite{Zavadlav:2014:ARS,Zavadlav:2014:ARSb,Zavadlav:2015:ARS}. In Section~\ref{secionathmosphere}, we have presented the results calculated with APBC using $n=1$ helical pitch. In particular, the simulated domain length is around 3.4{\,}nm long and considerably smaller than the DNA’s persistence length, which is about 50{\,}nm. To study mechanical properties of DNA, we need to increase the number of helical pitches as we have shown in Section~\ref{secperallatomMD} with our MD simulation results considering up to $n=10$ helical pitches along the $z$-direction of APBC simulation domain~(\ref{omegadomain}). 

To get further insight into the correct use of the APBC, we have started our investigation using a discrete worm-like chain (WLC) model in Section~\ref{WLCmodel}, where we have observed in Figure~\ref{figureWLC2} that the APBC affect less some local properties of the polymer chains than some global averages. In particular, the persistence length of the polymer chain can be estimated from local properties of relatively short polymer chains, simulated with the help of APBC. The APBC are also applicable to simulations of biopolymers with larger persistence length (for example, actin filaments~\cite{Floyd:2022:SBS,Gunaratne:2022:SBS2}), when a modeller is interested to understand the properties of the surrounding solvent.

In Appendix~\ref{appMD}, we provide the technical details of all-atom MD simulations, including the treatment of constant pressure simulations. The barostat used is again asymmetric with no fluctuations of $L_z$. In the APBC simulations, we have different treatment of the $z$-direction and all perpendicular directions in the $x-y$ plane. Simulations with 2D periodicity have also been used to study behaviour of a slab of water between two metallic walls~\cite{Hautman:1989:MDS}, which can be treated using three-dimensional Ewald techniques by including the image charges. One advantage of the APBC simulations is that they can be implemented with relatively minor modifications of standard all-atom MD tools~\cite{Humphrey:1996:VMD,Phillips:2020:NAMD,Li:2019:W3DNA,Lu:2008:3DNA} as detailed in Appendix~\ref{appMD}. Note that, the number of helical turns in the DNA model with APBC is fixed, and thus the model does not allow for over-winding or under-winding of DNA\cite{Marin-Gonzalez:2017:twist}.

\begin{acknowledgement}

This work was supported by the Engineering and Physical Sciences Research Council, grant
number EP/V047469/1, awarded to Radek Erban, and by the Japan Society for the Promotion of Science, KAKENHI grant number JP18KK0388 to Yuichi Togashi.
The computation was in part carried out using the computer resource offered under the category of General Projects by Research Institute for Information Technology, Kyushu University, and by the Advanced Research Computing (ARC) service at University of Oxford.

\end{acknowledgement}

\providecommand{\latin}[1]{#1}
\makeatletter
\providecommand{\doi}
  {\begingroup\let\do\@makeother\dospecials
  \catcode`\{=1 \catcode`\}=2 \doi@aux}
\providecommand{\doi@aux}[1]{\endgroup\texttt{#1}}
\makeatother
\providecommand*\mcitethebibliography{\thebibliography}
\csname @ifundefined\endcsname{endmcitethebibliography}
  {\let\endmcitethebibliography\endthebibliography}{}

\appendix

\section{Derivation of equation~(\ref{Ngenformula}) for the WLC model}

\label{appfjm}

Since $a_{\lfloor N/2 \rfloor} = a_1 = \ell$ for $N=2$ and $N=3$, equation~(\ref{Ngenformula}) is valid for $N=2$ and $N=3.$
Considering a general value of $N$, the constraint~(\ref{auxf1}) can be rewritten as
$$
\sum_{i=2}^N
\mathbf{l}_i
=
[0,0,L_z] 
-
\mathbf{l}_1
\, .
$$
Taking the scalar product of this equation with $\mathbf{l}_1$, we get
\begin{equation}
\sum_{i=2}^N
\left\langle
\mathbf{l}_1\cdot \mathbf{l}_i
\right\rangle
=
[0,0,L_z] 
\left\langle
\mathbf{l}_1
\right\rangle
-
\ell^2
=
\frac{L_z^2}{N} - \ell^2
\,,
\label{auxlhs}
\end{equation}
where we used
$
\left\langle
\mathbf{l}_1
\right\rangle
=
[0,0,L_z/N].
$
Assuming that $\left\langle
\mathbf{l}_1\cdot \mathbf{l}_i
\right\rangle$ are equal to each other on the left hand side of~(\ref{auxlhs}), we get
$
\left\langle
\mathbf{l}_1\cdot \mathbf{l}_i
\right\rangle
=
(L_z^2/N - \ell^2)/(N-1)
$
for $i=2,3,\dots,N$.
Substituting into (\ref{ajdef}),
we obtain
\begin{equation*}
a_{\lfloor N/2 \rfloor}
=
\displaystyle
\ell 
+
\frac{1}{\ell} \sum_{i=2}^{\lfloor N/2 \rfloor}
\left\langle
\mathbf{l}_1\cdot \mathbf{l}_i
\right\rangle
=
\frac{
\ell^2 N
(N-\lfloor N/2 \rfloor)
+
(\lfloor N/2 \rfloor - 1) L_z^2}{\ell N (N-1)}
\, ,
\end{equation*}
which is the equation~(\ref{Ngenformula}). Considering large values of $N$, equation~(\ref{Ngenformula}) simplifies to
$$
a_{\lfloor N/2 \rfloor}
\approx
\frac{\ell}{2}
+
\frac{L_z^2}{2 \ell N }
\, ,
$$
which can also be deduced from equation~(\ref{auxlhs}) by considering that the left hand side of equation~(\ref{auxlhs}) is twice the sum needed for the calculation of 
$a_{\lfloor N/2 \rfloor}.$

\section{All-atom MD simulation details}

\label{appMD}

Throughout the all-atom modeling and simulation, we used CHARMM36 force field, VMD 1.9.3 (for modeling)~\cite{Humphrey:1996:VMD}, and NAMD 2.14 (for simulation)~\cite{Phillips:2020:NAMD}.

\subsection{APBC implementation}

\label{appAPBC}

The periodic DNA models ($N$ bp) used for the APBC are generated as follows.
First, a $(N+1)$ bp long dsDNA configuration (in PDB format) is constructed with 3DNA (Web 3DNA 2.0~\cite{Li:2019:W3DNA}, in such a way that the $(N+1)$-th base pair is equivalent to the first base pair translated to the z-direction.
Here, we use the base pair step parameter set for B-DNA (calf thymus; generic sequence) with Twist 36.0 degrees and Rise 3.375 {\AA} for all nucleotide sequences.
Note that the resulting structure shows slight (sub-\aa{}ngstr\"om) shift between the first and the $(N+1)$-th (i.e. deleted) bases depending on the length and nucleotide sequence, which is then relaxed in the initial equilibration simulation run.
Then, a nucleotide at the 3'-end of each strand is removed, and the corresponding structure (in CHARMM PSF format) is generated by AutoPSF.
Finally, by editing the PSF file, the bond to the 3'-end (removed) nucleotide is substituted with that to the first base (i.e. 5'-end); and the angles and dihedrals (4 and 7 terms for each strand, respectively) are reconnected accordingly. These additional terms are listed in Table \ref{tab:psf}. Note that, for the CHARMM36 force-field, the type of the 5'-end phosphate (usually the first atom of each strand) should be changed from P to P2, as it is no longer at the terminal with the APBC.

For simulation with explicit solvent, each model has been solvated in a $L_{x} \times L_y \times L_z$ TIP3P water box where $L_{x} = L_{y} = 200$~\AA{} and $L_{z} = 3.375 N$~\AA, then neutralized by K$^{+}$ and 150 mM KCl has been added in the bulk.

\begin{table}[t]
    \centering
    \begin{tabular}{c c c c}
        \hline
        \multicolumn{4}{l}{Bonds} \\
        \hline
         $N$th O3' & 1st P \\
        \hline
        \hline
        \multicolumn{4}{l}{Angles} \\
        \hline
        $N$th C3' &	$N$th O3' & 1st P \\
        1st O1P & 1st P & $N$th O3' \\
        1st O2P & 1st P & $N$th O3' \\
        1st O5'	& 1st P & $N$th O3' \\
        \hline
        \hline
        \multicolumn{4}{l}{Dihedrals} \\
        \hline
        $N$th C4' &	$N$th C3' & $N$th O3' &	1st P \\
        $N$th C2' &	$N$th C3' & $N$th O3' &	1st P \\
        $N$th C3' &	$N$th O3' &	1st P &	1st O1P \\
        $N$th C3' &	$N$th O3' &	1st P &	1st O2P \\
        $N$th C3' &	$N$th O3' &	1st P &	1st O5' \\
        $N$th H3' & $N$th C3' & $N$th O3' &	1st P \\
        $N$th O3' &	1st P &	1st O5' & 1st C5' \\
        \hline
    \end{tabular}
    \caption{List of additional terms connecting the ends of each DNA strand.}
    \label{tab:psf}
\end{table}

\subsection{Molecular dynamics simulation}

After $10^4$ steps of energy minimization, each model was simulated for 10 ns at 2 fs time step (with SETTLE method for hydrogen).
Non-bonded interactions were cutoff at 12 {\AA} (with switching from 10 {\AA}), and particle mesh Ewald method was used for electrostatics.
Langevin thermostat at 300 K (damping: 1 per ps) and Nos\'{e}-Hoover Langevin piston barostat at 1~atm (period: 2~ps, decay: 1~ps) were applied. For the barostat, constant area setting was used, keeping $L_z$ (and $L_x$) constant and adjusting only $L_y$ to control the pressure.
The atomic positions were recorded at 10 ps intervals and used for analysis.

For simulations with periodic box length different values of $L_{z}$ in Figure~\ref{figure6}, the $z$-coordinate of each atom in the DNA model was rescaled, and then solvated into the water box with the adjusted value of $L_{z}$.

\subsection{Analysis}

The DNA structure in each snapshot was processed by 3DNA~\cite{Lu:2008:3DNA}. To consider also the base-pair step crossing the periodic boundary (i.e. the $N$-th to the first base pair), a periodic image of the first residue (shifted by $L_{z}$) was added at the end of each strand, and the resulting $(N+1)$ bp long segment was processed by 3DNA. The subsequent analysis then used the helical-axis positions and helical-axis vectors providing ${\mathbf h}_i$ in the average~(\ref{cosphij}). The distance between adjacent base-pair steps was obtained as the distance between the corresponding helical-axis positions, and other distances along the chain are calculated as the sum of those adjacent distances within the section. 
These values were averaged over three simulation trials except the initial 1 ns (Figures \ref{figure4}, \ref{figure5} and \ref{figure7}) or 5 ns (Figure \ref{figure6}) for each case.

Finally, the experimentally determined persistence length for our sequences of dinucleotides used in Figure~\ref{figure4} have been obtained in the literature~\cite{Geggier:2010:SDD} as
50.4~nm (poly(A)), 41.7~nm (poly(C)),  42.7~nm (poly(AT)), 49.6~nm (poly(CG)), 50.7~nm (poly(AC)) and 52.6~nm (poly(AG)).

\end{document}